\documentclass[preprint,showpacs,preprintnumbers,amsmath,amssymb]{revtex4}
\usepackage{epsfig,amsmath,amssymb,graphics,color,calc}

\newcommand{\be}{\begin{equation}}
\newcommand{\ee}{\end{equation}}
\newcommand{\ba}{\begin{eqnarray}}
\newcommand{\ea}{\end{eqnarray}}
\newcommand{\ta} {\tau_{\alpha}}
\renewcommand{\phi}{\varphi}

\begin{document}

\title{Probing the equilibrium dynamics of colloidal hard spheres
above the mode-coupling glass transition}

\author{G. Brambilla}
\affiliation{Laboratoire des Collo{\"\i}des, Verres
et Nanomat{\'e}riaux, UMR 5587, Universit{\'e} Montpellier II and CNRS,
34095 Montpellier, France}

\author{D. El Masri}
\affiliation{Laboratoire des Collo{\"\i}des, Verres
et Nanomat{\'e}riaux, UMR 5587, Universit{\'e} Montpellier II and CNRS,
34095 Montpellier, France}

\author{M. Pierno}
\affiliation{Laboratoire des Collo{\"\i}des, Verres et
Nanomat{\'e}riaux, UMR 5587, Universit{\'e} Montpellier II and CNRS,
34095 Montpellier, France}

\author{G. Petekidis}
\affiliation{IESL-FORTH and Department of Material Science and
Technology, University of Crete, GR-711 10 Heraklion, Greece}

\author{A. B. Schofield}
\affiliation{The School of Physics and Astronomy, Edinburgh
University, Mayfield Road, Edinburgh, EH9 3JZ, United Kingdom}

\author{L. Berthier}
\affiliation{Laboratoire des Collo{\"\i}des, Verres
et Nanomat{\'e}riaux, UMR 5587, Universit{\'e} Montpellier II and CNRS,
34095 Montpellier, France}

\author{L. Cipelletti}
\affiliation{Laboratoire des Collo{\"\i}des, Verres
et Nanomat{\'e}riaux, UMR 5587, Universit{\'e} Montpellier II and CNRS,
34095 Montpellier, France}

\date{\today}

\begin{abstract}
We use dynamic light scattering and computer simulations to study
equilibrium dynamics and dynamic heterogeneity in concentrated
suspensions of colloidal hard spheres. Our study covers an
unprecedented density range and spans seven decades in structural
relaxation time, $\ta$, including equilibrium measurements above
$\phi_{\rm c}$, the location of the glass transition deduced from
fitting our data to mode-coupling theory. Instead of falling out of
equilibrium, the system remains ergodic above $\phi_{\rm c}$ and
enters a new dynamical regime where $\ta$ increases with a
functional form that was not anticipated by previous experiments,
while the amplitude of dynamic heterogeneity grows slower than a
power law with $\ta$, as found in molecular glass-formers close to
the glass transition.
\end{abstract}




\maketitle

Hard sphere assemblies often constitute the simplest model to tackle
a variety of fundamental questions in science, from phase
transitions in condensed matter physics to the mathematics of
packing or optimization problems in computer science.
Experimentally, hard spheres systems are obtained using colloidal
particles~\cite{pusey}, emulsions, or granular
materials~\cite{jamming}. When crystallization is avoided, e.g. by
introducing size polydispersity, hard spheres at thermal equilibrium
become very viscous and eventually form an amorphous
solid~\cite{pusey2} at large volume fraction, $\phi$, in analogy to
the glass transition of molecular liquids~\cite{reviewnature} and
the jamming transition of grains~\cite{jamming}. However, the nature
of the colloidal glass transition, its precise location, the
functional form of the structural relaxation time divergence, and
the connection between slow dynamics and kinetic heterogeneities
remain largely open issues~\cite{vanmegen,chaikin}.

For hard spheres at thermal equilibrium, several distinct glass
transition scenarios have been described. In the first, the
viscosity or, equivalently, the timescale for structural relaxation,
$\tau_\alpha(\phi)$, diverges algebraically: \be \tau_\alpha (\phi)
\sim (\phi_{\rm c}- \phi)^{-\gamma}. \label{mct} \ee This is
predicted~\cite{mct} by mode coupling theory (MCT), and supported by
the largest set of light scattering data to date~\cite{vanmegen}.
Packing fractions $\phi_c \approx 0.57-0.59$ are the most often
quoted values for the location of the `colloidal glass transition'.
It is widely believed that a truly non-ergodic state is obtained at
larger $\phi$~\cite{pusey,pusey2,vanmegen}. Within MCT, the
amplitude of dynamic heterogeneity quantified by multi-point
correlation functions also diverges algebraically. In particular,
the four-point dynamic susceptibility should diverge as~\cite{JCP}:
$\chi_4 \sim (\phi_c-\phi)^{-2} \sim \ta^{2/\gamma}$, a prediction
that has not been tested experimentally.

Several alternative scenarios~\cite{free,ken,zamponi} suggest a
stronger divergence: \be \tau_\alpha (\phi) = \tau_{\infty}  \exp
\left[ \frac{A}{(\phi_{0} - \phi)^{\delta} } \right]. \label{vft}
\ee Equation (\ref{vft}) with $\delta=1$ is frequently used to
account for viscosity data~\cite{chaikin} because it resembles the
Vogel-Fulcher-Tammann (VFT) form used to fit the viscosity of
molecular glass-formers~\cite{reviewnature}, with temperature
replaced by $\phi$. Moreover, it is theoretically expected on the
basis of free volume arguments~\cite{free}, which lead to the
identification $\phi_0 \equiv \phi_{\rm rcp}$, the random close
packing fraction where osmotic pressure diverges. Kinetic arrest
must occur at $\phi_{\rm rcp}$  (possibly with $\delta = 2$~\cite{ken}),
because all particles block each other at that
density~\cite{pointJ,ken,torquato2}. Entropy-based theories and
replica calculations~\cite{zamponi} predict instead a divergence of
$\ta$ at an ideal glass transition at $\phi_0 < \phi_{\rm rcp}$,
where the configurational entropy vanishes but the pressure is still
finite. Here, the connection to dynamical properties is made through
nucleation arguments~\cite{wolynes} yielding Eq.~(\ref{vft}), with
$\delta$ not necessarily equal to unity~\cite{BB}. In this context,
the amplitude of dynamic heterogeneity should increase only
moderately, typically logarithmically slowly in
$\ta$~\cite{toninelli}.

In molecular glass-formers where dynamical slowing down can be
followed over as many as 15 decades, the transition from an MCT
regime, Eq.~(\ref{mct}), to an activated one, Eq.~(\ref{vft}), has
been experimentally demonstrated~\cite{reviewnature}. For colloidal
hard spheres, the situation remains controversial, because dynamic
data are available over a much smaller
range~\cite{pusey,vanmegen,chaikin,Chaikin2}, typically five decades
or less. Crucially, equilibrium
measurements  were reported only for $\phi <
\phi_\mathrm{c}$,
leaving unknown the precise nature and location of
the divergence. Theoretical claims exist that the cutoff mechanism
suppressing the MCT divergence in molecular systems is inefficient
in colloids due to the Brownian nature of the microscopic dynamics,
suggesting that MCT could be virtually exact~\cite{brownian}. This
viewpoint is challenged by more recent MCT calculations~\cite{ABL},
and by computer studies of simple model systems where MCT
transitions are avoided both for stochastic and Newtonian
dynamics~\cite{stochastic1,stochastic2}.

Here, we settle several of the above issues by studying the
equilibrium dynamics of colloidal hard spheres using dynamic light
scattering and computer simulations. By extending previous data by
at least two orders of magnitude in $\ta$, we establish that the
volume fraction dependence of both $\ta$ and $\chi_4$ follows MCT
predictions only in a restricted density range below our fitted
$\phi_{\mathrm{c}} \approx 0.59$. Unlike previous studies, we
provide equilibrium measurements \textit{above} $\phi_{\mathrm{c}}$,
thereby proving unambiguously that in our sample the algebraic
divergence at $\phi_\mathrm{c}$ is absent. Instead, a new regime is
entered at larger $\phi$, where the dynamics is well described by
Eq.~(\ref{vft}) with $\delta \approx 2$ and $\phi_0$ much larger
than $\phi_{\mathrm{c}}$. The amplitude of kinetic heterogeneities
then grows slower than a power law with $\ta$, as in molecular
glasses close to the glass transition.

Dynamic light scattering (DLS) experiments are performed in the
range $0.01 < \phi < 0.5981$. We use poly-(methyl methacrylate)
(PMMA) particles of average diameter $\sigma = 260$~nm, stabilized
by a thin layer of grafted poly-(12-hydroxy stearic acid) (PHSA).
The size polydispersity, about $10\%$, is large enough to prevent
crystallization on a timescale of at least several months. The
particles are suspended in a mixture of cis-decalin and tetralin
that almost perfectly matches their average refractive index,
allowing the dynamics to be probed by DLS. Additionally, a careful
analysis of the combined effects of optical and size polydispersity
shows that we probe essentially the self-part of the intermediate
scattering function~\cite{R1}: \be F_s(q,t) = \left\langle
\frac{1}{N} \sum_{j=1}^N e^{i {\bf q} \cdot ({\bf r}_j(t)-{\bf
r}_j(0))} \right\rangle. \label{fs} \ee Here, ${\bf r}_j(t)$ is the
position of particle $j$ at time $t$, $\mathbf{q}$ is the scattering
vector ($q \sigma=6.5$, close to the first peak of the static
structure factor) and brackets indicate an ensemble average. A
combination of traditional~\cite{Berne1976} and
multispeckle~\cite{Sillescu} DLS is used to measure the full decay
of $F_s(q,t)$. We carefully check equilibration by following the
evolution of the dynamics after initialization, until $F_s$ stops
changing over a time window of at least $10 \tau_{\alpha}$. Samples
are prepared by dilution, starting from a very concentrated batch
obtained by centrifugation. All volume fractions relative to that of
the initial batch are obtained with a relative accuracy of
$10^{-4}$, using an analytical balance and literature values for
particle and solvent densities~\cite{Chaikin2}. Relative volume
fractions are converted to absolute ones by comparing the
experimental $\phi$ dependence of the short-time self-diffusion
coefficient measured by DLS to two sets of theoretical
calculations~\cite{R2} at low density, $\phi \le 0.2$. For less
polydisperse samples, this calibration method yields $\phi$ values
compatible with those obtained by mapping the experimental freezing
fraction to $\phi_{\rm f}=0.494$~\cite{SegrePRE}. The remaining
uncertainty on the absolute $\phi$ is about 5\%, because \cite{R2}
contains two slightly different predictions. To ease the comparison
with the simulations, we set the absolute $\phi$, within this
uncertainty range, so that our experimental and numerical
$\tau_{\alpha}$ closely overlap for $\phi
> 0.55$.

We use a standard Monte Carlo algorithm~\cite{stochastic2} to study
numerically a 50:50 binary mixture of hard spheres of diameter
$\sigma$ and $1.4 \sigma$, known to efficiently prevent
crystallization. We work in a three dimensional space with periodic
boundary conditions, and mainly use $N=10^3$ particles. No
noticeable finite size effects were found in runs with
$N=8\cdot10^3$ particles performed for selected state points. In an
elementary move, a particle is chosen at random and assigned a
random displacement drawn within a cubic box of linear size
$0.1\sigma$ centered around the origin. The move is accepted if the
hard sphere constraint remains satisfied. One Monte Carlo step
corresponds to $N$ such attempts. The dynamics is characterized by
the self-intermediate scattering function, Eq.~(\ref{fs}), measured
for $q \sigma = 6.1$, close to the first diffraction peak.

\begin{figure}
\psfig{file=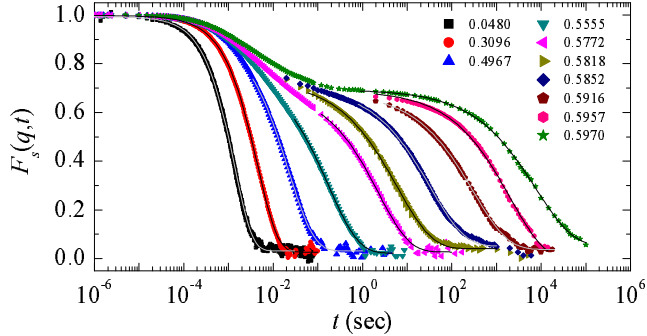,width=8.5cm} \caption{\label{fig1} (Color
online) Time dependence of the self-intermediate scattering function
$F_s(q,t)$ in DLS experiments at $q \sigma = 6.5$ for representative
volume fractions. Lines are stretched exponential fits to the final
decay, yielding relaxation times spanning about 7 decades.
Ergodicity is preserved above the (avoided) MCT glass transition at
$\phi_c \approx 0.59$.}
\end{figure}

Representative $F_s(q,t)$ obtained by DLS are plotted in
Fig.~\ref{fig1}, showing that the relaxation is fast and
monoexponential at low $\phi$, while a two-step decay is observed
when increasing $\phi$, reflecting the increasingly caged motion of
particles in dense suspensions~\cite{pusey2}.
We measure the structural relaxation
time by fitting the final decay of $F_s$ to a stretched exponential,
$F_s(q,t) = B \exp[-(t/\ta)^{\beta}]$.

\begin{figure}
\psfig{file=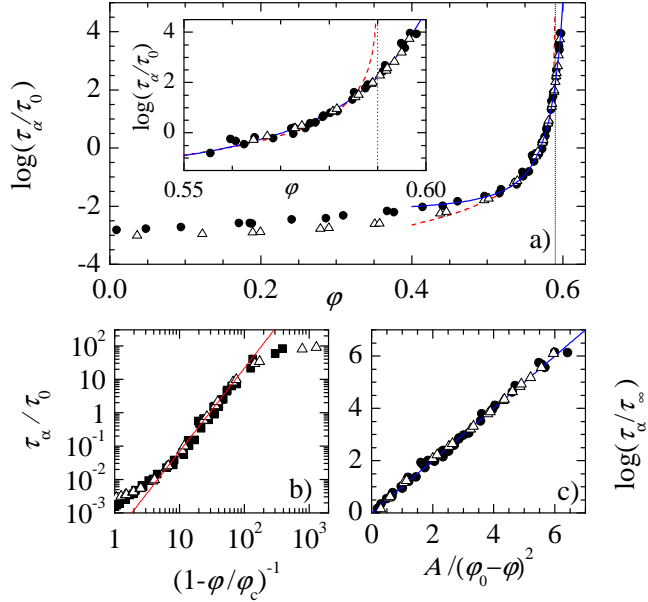,width=8.5 cm} \caption{\label{fig2} (Color
online) a) Relaxation timescale $\tau_{\alpha}$ for hard spheres in
experiments (black circles) and simulations (open triangles),
respectively in units of $\tau_0 = 1$ sec and $\tau_0 = 7 \cdot
10^4$ MC steps. The red dashed line is a power law fit,
Eq.~(\ref{mct}), with $\phi_{\mathrm{c}} = 0.590$ (vertical dotted
line) and $\gamma=2.5 \pm 0.1$. The continuous blue line is a fit to
DLS data using Eq.~(\ref{vft}), with $\phi_0=0.637$ and $\delta=2$.
The zoom in the inset shows that the MCT singularity is absent. b)
Same data plotted against $1/(1-\phi/\phi_c)$. A straight line with
slope $\gamma$ is obtained in an MCT regime covering almost 3
decades in $\ta$. c) Data for $\phi > 0.41$ plotted using reduced variables
with $\phi_0 = 0.637$ and $0.641$ for experiments and simulations,
respectively.}
\end{figure}

Figure~\ref{fig2}a shows $\ta(\phi)$ for both experiments and
simulations. Time units are adjusted to maximize the overlap
($\approx$ 5.5 decades) of both data sets at high $\phi$. Our
experimental data are well fitted by Eq.~(\ref{mct}) in the range
$0.49 < \phi \le 0.585$, with $\phi_{c} = 0.590 \pm 0.005$ and
$\gamma= 2.5 \pm 0.1$. For the slightly less polydisperse sample of
Ref.~\cite{vanmegen}, a similar power law behavior with $\gamma
\approx 2.7$ and $\phi_c=0.571-0.595$ was reported, the two quoted
values of $\phi_c$ stemming from experimental uncertainty in the
volume fraction determination. However, our measurements for the
largest densities strongly deviate from the MCT fit. Attempts to
include points at $\phi > 0.59$ in the MCT fit yield unphysically
large values of $\gamma$. Similar deviations are found in our
simulations, showing that hydrodynamic interactions play little role
in experiments performed at large $\phi$, although they probably
explain the discrepancy with simulations at low volume fraction, see
Fig.~\ref{fig2}a. Therefore, our results unambiguously demonstrate
that the mode-coupling singularity is absent in our hard sphere
colloidal system, as is also found in molecular
glass-formers~\cite{reviewnature}.

What is the fate of the fluid phase above $\phi_c$?
Figures~\ref{fig2}a and \ref{fig2}c show that the increase of $\ta$
at high $\phi$ is extremely well described by an exponential
divergence, Eq.~(\ref{vft}). We find that the data can be fitted
well using the conventional form with $\delta = 1$, yielding
$\phi_0(\delta=1) \approx 0.614 \pm 0.002$. This is consistent with
previous analysis of viscosity data~\cite{chaikin}.
However, the quality of the fit improves when the exponent $\delta$
is allowed to depart from unity. The optimal value, robust for both
experimental and numerical data, is $\delta  = 2.0 \pm 0.2$, which
yields our best estimate for the location of the dynamic glass
transition: $\phi_0 \approx 0.637 \pm 0.002$ (experiments) and
$\phi_0 \approx 0.641 \pm 0.002$ (simulations). Figure~\ref{fig2}c
shows the linear dependence of $\log \ta$ on $(\phi_0-\phi)^{-2}$,
demonstrating the exponential nature of the dynamic singularity.

\begin{figure}
\psfig{file=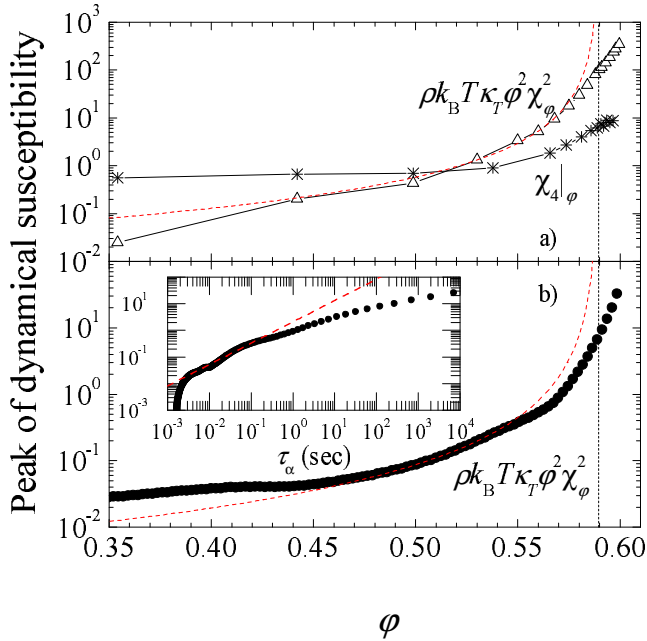,width=8.5 cm} \caption{\label{fig3} (Color
online) Peak of dynamic susceptibilites, Eq.~(\ref{ensemble}),
measured in a) simulations and b) experiments. In a) both
contributions to $\chi_4$ are compared, validating $\chi_\phi$ as a
valuable tool to quantify dynamic heterogeneity in hard spheres. The
predicted MCT algrebraic divergence (red dashed line) holds over a
small density range. The inset shows that the size of dynamic
heterogeneities grows slower than a power law at large $\ta$, as
found in molecular glass-formers.}
\end{figure}

The behavior of dynamical heterogeneity provides additional evidence
of a crossover from a restricted MCT regime to an `activated' type
of dynamics. Using methods detailed in \cite{science,dalleferrier},
we study the evolution of the three-point dynamic susceptibility
defined by: $\chi_\phi(q,t) \equiv
\partial F_s(q,t) / \partial \phi$.
This linear response function is directly connected to a four-point
dynamic susceptibility: $\chi_4(q,t) = N \langle \delta F_s(q,t)^2
\rangle$, where $\delta F_s(q,t)$ denotes the fluctuating part of
the self-intermediate function; $\chi_4$ is a powerful tool to
quantify dynamic heterogeneity in glass-formers~\cite{toninelli},
because it represents the average number of molecules whose dynamics
are correlated. In hard spheres, the following relation
holds~\cite{science}: \be \chi_4(q,t) = \chi_4(q,t)|_\phi + \rho k_B
T \kappa_T (\phi \chi_\phi(q,t))^2, \label{ensemble} \ee where
$\rho$ is the number density, $\kappa_T$ the isothermal
compressibility (measured in simulations, taken from the
Carnahan-Starling equation of state in experiments), and
$\chi_4(q,t)|_\phi$ denotes the value taken by $\chi_4(q,t)$ in a
system where density is strictly fixed. Only the second contribution
to $\chi_4(q,t)$ in (\ref{ensemble}) can be accessed experimentally,
but both terms can be determined in simulations. We obtain
$\chi_\phi(q,t)$ by applying the chain rule to the fitted
$F_{\mathrm{s}}(B,\ta,\beta)$~\cite{dalleferrier}, where the $\phi$
dependence of $B,\ta,\beta$ is fitted by smooth polynomials. Our
results are independent of the choice of fitting functions, and
consistent with that obtained from finite differences between data
at nearby $\varphi$, when available. Figure~\ref{fig3} shows the
peak of dynamical susceptibilities as a function of $\phi$. First,
we numerically establish in Fig.~\ref{fig3}a that the term
comprising $\chi_\phi$ is the main contribution to $\chi_4$ when
$\phi > 0.52$, implying that a good estimate of $\chi_4$ can be
obtained using three-point functions in hard spheres, as surmised in
\cite{science}, and established for molecular glass-formers in
\cite{JCP}. For both simulations and experiments, the MCT prediction
for the algebraic divergence of $\chi_4(q,t)$ only holds over a
limited density range. Indeed, when plotted against $\ta$, $\chi_4$
eventually grows slower than a power law, as found for the size of
dynamically correlated regions in molecular glasses close to the
glass transition~\cite{dalleferrier}--a hallmark of activated
dynamics~\cite{toninelli}.

Our results establish the existence of a non-trivial, exponential
divergence of $\ta(\phi)$ at a critical volume fraction $\phi_0
\approx 0.637$ much above the putative `colloidal glass transition'
at $\phi_{\mathrm{c}} \approx 0.59$. It is natural to ask whether
$\phi_0$ and $\phi_{\rm rcp}$ coincide. This is a difficult question
because $\phi_{\rm rcp}$ can always be shifted to a larger value by
trading order and packing~\cite{torquato2}. For the binary mixture
studied here, the onset of jamming has been located at $\phi_J =
0.648$~\cite{pointJ}. Furthermore, for 10~\% polydispersity, the
estimate $\phi_{\rm rcp} \approx 0.67$ was obtained in numerical
work~\cite{simusillescu}, well above $\phi_0$. Finally, we have
employed Monte Carlo simulations to produce disordered hard sphere
configurations with finite pressure above $\phi_0$ by a very fast
compression of fluid configurations used to produce the equilibrium
data in Fig.~\ref{fig1} (open symbols). These results support the
possibility that $\phi_0 < \phi_{\rm rcp}$, implying a fundamental
difference~\cite{jorge} between the glass~\cite{zamponi} and
jamming~\cite{pointJ} transitions in hard spheres.

In conclusion, we report a set of dynamic data for a well-known
colloidal hard sphere system covering an unprecedented dynamic range
of equilibrium relaxation timescale. While the onset of dynamical
slowing can be described by an MCT divergence at a critical volume
fraction $\phi_\mathrm{c}$, upon further compression a crossover
from an algebraic to an exponential divergence at a much larger
volume fraction $\phi_0$ is observed, accompanied by a similar
crossover for the growth of dynamical correlations. Our results show
that the apparent singularity at $\phi_\mathrm{c}$ does not
correspond to a genuine `colloidal glass transition', suggesting
that the MCT transition is generally avoided in colloidal materials,
just as in molecular glass-formers.

This work was supported by the Joint Theory Institute at the
University of Chicago, the European MCRTN ``Arrested matter''
(MRTN-CT-2003-504712), the NoE ``SoftComp'' (NMP3-CT-2004-502235),
Region Languedoc-Roussillon and ANR ``DynHet'' and CNES grants. L.C.
acknowledges support from the IUF.

\end{document}